# Pivoting on the spatial dimension for mobile telecom energy-saving

Liyan Xu[1#*], Yihang Li[1,2#], Hang Yin[1], Hongbin Yu[1], Yuqiao Deng[1], Qian Huang[3*], Yinsheng Zhou[3], Yu Liu[4*]

## Author Information

[1] College of Architecture and Landscape Architecture, Peking University, Beijing 100871, China.

[2] College of Urban and Environmental Sciences, Peking University, Beijing 100871, China.

[3] Service Lab, Huawei Technologies. Beijing, 100194, China.

[4] Institute of Remote Sensing and Geographical Information Systems, Peking University, Beijing 100871, China.

## Contributions

Liyan Xu, Yihang Li, and Qian Huang conceptualized this study and contributed to the interpretation, manuscript writing, and development. Yihang Li, Hang Yin, Hongbin Yu, and Yuqiao Deng carried out the data processing and analysis. Yinsheng Zhou contributed to the data acquisition and preprocessing. Yu Liu and Liyan Xu supervised the study.

[#] These authors contributed equally in the paper.

## Corresponding authors

Correspondence to: Liyan Xu, Qian Huang, and Yu Liu.

## Abstract

Improving energy efficiency is vital for the 5G mobile telecommunication network, while the conventional temporal shutdown strategy is exhausting its energy-saving potential. Inspired by the symmetricity between the temporal peaks and valleys and the geographical hot and cold-spots in network traffic, we propose a new strategy for telecom energy-saving which matches the base station cells with traffic hotspots as much as possible to reduce waste. We showed deductively and empirically that the probability density function of the spatial distribution of network traffic typically has a steep power-law form, which gives rise to temporally stable traffic hotspot areas that occupies only about 13% of the space and also have proper sizes suitable for spatial optimization. A simplified model hence built yields up to 19.59% energy savings for the 4G network, and 29.68% for a hypothetical 5G network, prominent margins over what the temporal shutdown approach would gain.

Reducing carbon emission through conserving fossil energy consumption[1] has been one of the major strategies adopted by countries around the world in response to global climate change, and its importance is particularly evident today when energy prices are soaring under the duress of various uncertainties [2,3]. The power consumption of the world's massive mobile communication network (MCN) constitutes a significant part of energy use [4]. In the 4G era, the electricity costs of base stations and their ancillary facilities account for more than 30% of the enterprise' operating expenses[5], which is already a heavy burden. However, the percentage is expected to double in the 5G era [6]. This is because the 5G network utillizes high-frequency bands to enable high information capacity, which results in a much lower service coverage radius of a single base station (maximum 100-150m) [7] than that of a 4G base station (about 1-3km) [8], and consequently requires more base stations. Meanwhile, the power of a typical 5G base station is usually 2-3 times or even higher than that of a 4G one [4,9]. A simple calculation, therefore, shows that the overall energy consumption of the 5G network is expected to be about one order of magnitude higher than that of 4G.

In the face of the high energy consumption of the MCN, people have been actively pursuing energy-saving approaches. Most of the existing strategies focus on the temporal dimension, and achieve energy saving by shuting down inactive or low-trafiic load base stations during non-peak times [10,11]. However, this strategy has a major limitation. Let us consider the relationship between a base station's power ($P$) and its traffic load ($T$), which is approximately a linear function under normal load with the slope $a$ [12,13]:

$$P = aT + b(R^\alpha) \quad (1)$$

Where the intercept term $b(R)$ reflects the no-load power of a base station, including the power of the baseband and air conditioning facilities, ete., and it is a positive function of the distance between the base station and the traffic users ($R$) in which $\alpha$ is the signal propagation path attenuation coefficient which is a positive costant. The equation illustrates a fundamental dilemma facing MCN energy efficiency: on the one hand, as the no-load power always exists regardless whether there is traffic, one cannot save the corresponding power consumption through carrier and symbol-level shutdown operations [15]. On the other hand, cell-level shutdown, which closes the base station cell completely, although can save the power consumption, will render the entire service area signaless, and is thus inpractical in most occasions [16,17]. Overall, the existence of no-load power during non-peak times constitutes a major limitation in the improvement of MCN energy efficiency.

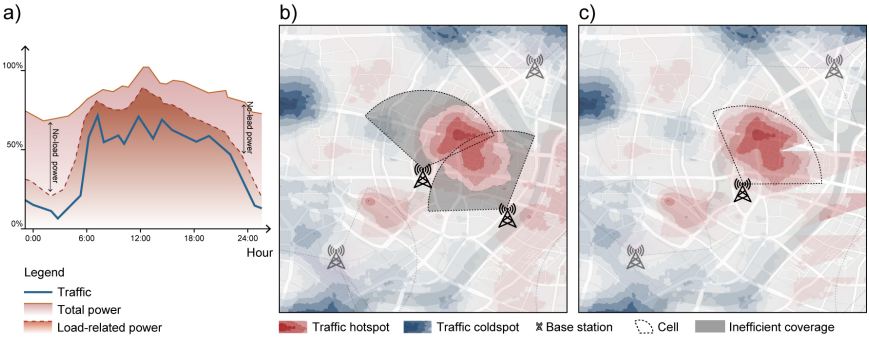

**Fig. 1 Spatial-temporal symmetry in telecom resource demand and energy-saving potential pivoting on the spatial dimension**

*a Uneven distribution of traffic along time forms peaks and valleys; shutdown operations in non-peak time help save energy;*

*b Uneven distribution of traffic in space forms hot and cold-spots; base station cells covering both traffic hot and cold-spots render waste of energy;*

*c Spatially optimized base stations with cells as dedicated to traffic hotspots as possible yield energy savings.*

Therefore, the key to further reduction of MCN energy consumption is to reduce the number of base stations with inevitable no-load power as much as possible without hampering signal coverage. On this, the apparent analogy between the temporal peaks/valleys and spatial hot/cold spots in MCN traffic inspires the possibility of an energy-saving strategy stemming from the *spatial* dimension. To explain the strategy, we first introduce the division of the two different functional layers of base stations in an MCN which is the common practice in network planning, namely the coverage and capacity layers, with the former providing basic service coverage and the latter satisfying additional traffic demand [18] [19]. Hence, despite that the no-load power of the coverage-layer base stations is unavoidable, since the spatial distribution of traffic is heterogeneous, one may achieve energy saving through optimizing the spatial distribution of the capacity-layer base stations such that their service coverage is as dedicated to high-traffic areas as possible. As illustrated in Fig. 1(b) and (c), on the one hand, by minimizing the "wasted" MCN resources, i.e., those dedicated to low-traffic areas, one may use as less base stations as possible to satisfy the traffic demand, and thus energy is saved from the reduced numbers of base stations. On the other hand, one may also perform cell-level shutdown on such dedicated capacity-layer base stations at non-peak times to further save energy without worrying induced service deterioration elsewhere. Besides, in Fig. 1(c), such placed base stations would be closer to the users than in Fig. 1(b), i.e., the service range *R* in Equation (1) is smaller, the no-load power is consequently smaller. Either way, the spatial optimization approach helps match the demand and supply of MCN resouces spatially, and hence improves energy efficiency by reducing wastes.

Despite the tempting vision above, it requires the spatio-temporal distribution of MCN traffic to satisfy three conditions: 1) traffic should be distributed heterogeneously enough, or putting another way, rather skewed such that the traffic hotspot areas only occupy a small portion of the entire service area and hence demand a small number of capacity-layer base stations; 2) the spatial distribution of traffic possesses high spatial autocorrelation with proper characteristic scales, i.e., high traffic areas tend to form continuous clusters in space of which the size matches that of the service coverage radius of base stations. The condition is necessary because the shutdown of base stations can only be performed in the spatial unit of a cell, and therefore the scale of the high traffic areas must match with the characteristic scale of the cell's service coverage radius. Otherwise, if the former are so spatially dispersed that each of them occupies only a fraction of a cell, the rest of the cell will fall in low-traffic areas, which contridictes the spatial optimization strategy; and 3) such spatial structure is stable over time. This is because for operational concerns, we wish the spatial optimization strategy to be effective over a long enough time period such that high-frequent adjustment of the base station locations, which would be inpractical, is not necessary.

We use empirical data to test whether these conditions hold, to which our hypotheses are all yes. The empirical MCN traffic data are from Taiyuan, Shanxi Province, China (see the Methods section for data description). The case is chosen because being a mid-large sized city of about four million city proper population and five million metro area population, Taiyuan is a rather "ordinary" city as a mobile communication market, and thus is of typical value for our analysis. After testing the hypothese, we also develop a simple spatial optimization model, and show with a set of numerical cases the potentially large energy-saving space of the spatial optimization strategy under typical 4G and 5G network cirmumstances.

## Results

### The highly skewed spatial distribution of the MCN traffic

The spatial distribution characteristics of the MCN traffic, to our knowledge, is rarely addressed in the existing literature. Nevertheless, the spatial distribution of certerin closely related physical quantities, i.e., mobile signaling [20] and population [21,22], which is the user of the MCN, exhibit highly skewed patterns. These facts inspire one to hypothesize that the spatial distribution of the traffic should exhibit similar skewed patterns. Indeed, if the skewness is significant, it intuitively implies that high-traffic areas would account for only a small fraction of the space, thus suggesting considerable energy saving potential. And technically, such potential would be realizable given the current "hybrid networking" [4] strategy in 5G network building.

For a conceptual deduction of the hypothesis, based on the maximum entropy principle [23,24] which is found prevalent in human behavior, the generalized form of the spatial distribution function $P(X)$ of traffic can be derived as follows (see the Methods section for the detailed deduction):

$$P(X = k) = e^{\lambda_0 + \lambda_1 \cdot c(k) - 1} \tag{2}$$

Where $P(X = k)$ is the probability function when the random variable $X$ for the volume of the traffic (in bits) takes the value $k$; $c(k)$ denotes the number of people (users) in the spatial location where $X = k$; the parameters $\lambda_0$ and $\lambda_1$ are Lagrangian multipliers that ensure the maximization meets these constraints (see the Methods section for the detailed deduction); negative number -1 is from the solution of the Lagrangian function. Overall, this function takes a generalized exponential form, and in the extreme case when the implicit function $c(k)$ takes a logarithmic form, it may be reduced to a power-law function, while other possible forms of $c(k)$ will lead to more skewed distributions (see the Methods section for details). This means that the spatial distribution of the traffic is indeed likely to be significantly skewed.

We use the traffic data from Taiyuan, China to empirically test the hypothesis. By geographical laws, the spatial distribution of traffic is potentially influenced by the spatial scope [25] and context [26] of analysis. Therefore, we summarized the probability density function (pdf) of the spatial distribution of traffic (daily aggregated) at two granularities: 50 * 50m and 500 * 500m, and at four spatial scales: citywide (~300 km²), district (~ 50 km²), neighborhood (~ 20 km²), and urban place (about 0.2 km²), respectively (Table 1). Results show that: 1) the traffic exhibits a power-law distribution at the citywide scale; as the scale of the analysis going downward, the distribution remains in power-law form in most cases, although in other cases it shows an exponential distribution, or in rare cases a log-normal distribution. 2) At all scales and the 50 * 50m resolution, the $\alpha$ parameters of the power-law distribution form a normal distribution with a mean of 2.3-2.5 and a standard deviation of about 0.36-0.45; in other words, the power-law distributions here are very "steep", and it appears that the smaller the scale is, the steeper the distribution becomes. In the case of the exponential and log-normal distributions, they are even more highly skewed. 3) The above distribution forms appear to be insensitive to the granularty of analysis. At the spatial resolution of 500 * 500m, the mean of $\alpha$ of the power-law distribution is slightly reduced to about 2, and the standard deviation increases to 0.46-0.6, i.e., it still maintains a pattern very similar to that at the 50 * 50m resolution; and the situation is also similar in the case of the exponential and log-normal distributions. Given the above-stated service coverage radiuses of the network, the result indicates that the distributional skewness is meaningful for our energy-saving strategy for both the 4G and 5G networks.

**Table 1 Summary of the pdf functions and parameters of the spatial distribution of MCN traffic at multiple spatial scales and granularities**

| pdf of MCN traffic | Spatial scope | | | |
|---|---|---|---|---|
| | Citywide (1 Unit of Analysis) | District (6 Units of Analysis) | Neighborhood (12 Units of Analysis) | Urban Place (1228 Units of Analysis) |
| Granularity = 500 * 500m | Power-law $\alpha = 2.5235$ | Power-law (66.6%) $\mu_\alpha = 2.3103$ $\sigma_\alpha = 0.4447$ others: Exponential (33.4%) | Power-law (53.84%) $\mu_\alpha = 2.3979$ $\sigma_\alpha = 0.3635$ others: Exponential (38.46%); Log-normal (7.69%) | Power-law (45.78%) $\mu_\alpha = 2.3661$ $\sigma_\alpha = 0.4377$ others: Exponential (54.22%) |
| Granularity = 50 * 50m | Power-law $\alpha = 2.5992$ | Power-law (66.6%) $\mu_\alpha = 2.0573$ $\sigma_\alpha = 0.5456$ | Power-law (61.5%) $\mu_\alpha = 1.9532$ $\sigma_\alpha = 0.4687$ | Power-law (49.3%) $\mu_\alpha = 2.1051$ $\sigma_\alpha = 0.5982$ |

|   |   |   |   |
|---|---|---|---|
|   | others: | others: | others: |
|   | Exponential | Exponential | Exponential |
|   | (33.4%) | (38.5%) | (50.7%) |

## Existence of stable MCN traffic hotspot areas

Next, we test the second and third hypotheses. For the spatial autocorrelation hypothesis, calculations show that in the previous case, traffic exhibits significant spatial autocorrelation, with a global Moran's I of 0.8265 (Fig. 2a) [27]. Further, we define the minimal convex hull of high traffic locations as the traffic hotspot area, and through the LISA (Local Indicator of Spatial Association) index [28] which gives the "high-high" autocorrelation regions, and subsequent spatial clustering using DBscan[29], we find in our case that the detected traffic hotspot areas (Fig. 2b) do meet the spatial scale requirements with the smallest hotspot area covering 5 ha$^2$, an area size that is bigger than a typical cell (Fig. 2c). For the temporal stability hypothesis, it should be noted that here temporal stability does not require the absolute stability of traffic in each cell, which is obviously impossible. Instead, what matters is the relative traffic ranking among the cells along time, i.e.,whether the traffic hotspot areas at one moment still be the ones at the next. A rank clock analysis [30] which yields the ranking of hourly traffic of all cells during a week revelas that among the 168 cells in traffic hotspot areas, more than 150 of them ranked high in all time of the week (occupying the central part of the rank clock). Specifically, although their absolute ranking fluctuated at different times, more than 80% of the traffic hotspot cells are consistently among the top 168 during more than 80% of the time, with the remaining 20% of the time periods majorly between 3:00 and 6:00 a.m. when traffic usage is low and has no significant energy savings challenges anyway. Similarly, the traffic coldspots, although also experiencing fluctuations in ranking, remained coldspots at all times. Moreover, we observe in the rank clock that there seems to be an insurmountable gap between the hotspot and the coldspot cells, implying that the traffic hotspot areas are indeed stable over time (Fig. 2d). Overall, the two conditions combined give rise to the existence of stable traffic hotspot areas, which in our case only account for 13.08% of the total study area while capturing 63.68% of the total traffic (Fig. 2c).

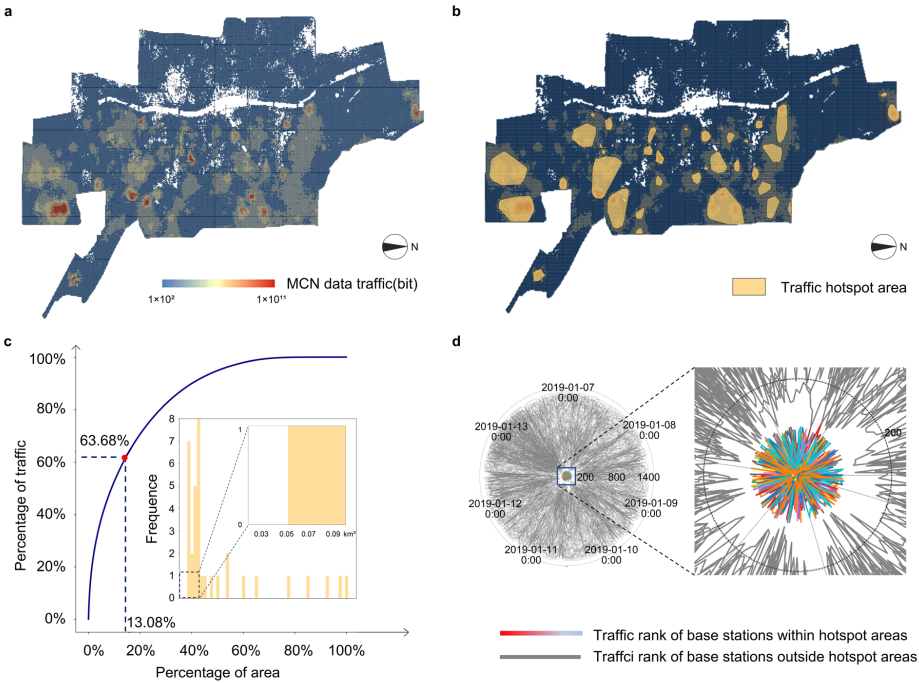

**Fig. 2 The spatiotemporal distribution of MCN traffic in Taiyuan**

*a Spatial distribution of traffic;*

*b Traffic hotspot areas based on spatial cluster of high-high locally autocorrelated locations;*

*c Cdf of spatial distribution of traffic, where all traffic hotspot areas only account for 13.08% of the total study area while capturing 63.68% of the total traffic; small window shows a histogram of traffic hotspot area sizes, where one can see that the smallest traffic hotspot has an area size greater than 0.05 $km^2$;*

*d Rank clock of traffic within a week, where the first-ranked traffic location is in the center of the clock and the lower-ranked locations outwwards. The gray lines indicate the traffic coldspot and the colored lines the traffic hotspot locations.*

## Spatial optimization through supply-demand coupling of MCN resources for energy-saving

Finally, we estimate the potential effect of an MCN energy-saving strategy through spatial optimization with the same case. As analyzed earlier, this is achieved by spatially coupling the supply and demand of MCN resources, such that more base stations are allocated for high traffic areas as the capacity layer of the cellular network, and less for the traffic coldspots only to ensure basic service coverage. In this picture, ideally the pdf of the traffic distribution should show a single-peak pattern, with fewer base stations covering both very low and very high traffic areas at the head (low-traffic) and the tail (high-traffic) of the distribution, and more in the middle part. This is because, if the probability density of the head was large, one can reduce energy consumption by shutting down the base stations with very low loads and shifting their loads to others nearby; while at the tail of the distribution, we know from previous empirical findings that they only occupy a small portion of the city space, and therefore only a small number of base stations should serve. Fig. 3(c) shows the status-quo traffic distribution at the cell level, and one can see that its pdf follows an exponential distribution. In other words, there exists a large number of low-load base stations, and they are undoubtedly the source of low energy efficiency. The supply-demand coupling of MCN resources, i.e., equalizing the spatial distribution of base stations according to the spatial distribution of traffic, should make the two match with each other.

In estimating the energy-saving potential of such optimization, a challenge is that while the citywide spatial distribution of traffic obeys a steep power-law distribution, the pattern cannot be directly extrapolated to smaller scales due to geographical heterogeneity (Table 1). Hence, we cannot simply follow the pattern of power-law distribution at all spatial scales to calculate the proportion of high traffic areas and, consequently, estimate the energy-saving potential. Therefore, we propose a simplified spatial optimization model for energy savings estimation. The model has an objective of minimizing the total power of all base stations $\sum_i P_i$. With the location of base stations as the decision variable, it optimizes the spatial arrangement of base stations such that the supply of MCN resources matches the geographical pattern of traffic demand. Hence, the model can be technically transformed as a modified maximum set-cover problem [31]: the traffic coverage $\sum_{i \in I} y_i$ is maximized with a proper set of base station locations, which is subject to a series of constraints regarding the service coverage radius of the base stations, signal strength, along with other technical constraints (See the Methods section for details of the optimization model).

We design four basic scenarios to conduct the experiment, which varies in 1) whether the number of base stations is included as a decision variable besides the locations of base stations, and 2) whether the optimization is performed for the entire study area or just in the traffic hotspot areas. All above scenarios use the power parameters of a typical 4G (LTE 2T2R) base station [32], and use the status-quo power as the baseline for energy-saving effect calculation. Additionally, we conduct another experiment for a hypothetical 5G scenario, in which major changes involve the service coverage radius and power of individual base stations. We set the former to 150 m and the latter according to literature [16]. Also, since the 5G scenario is a counter-fact and there is no real-world baseline, we refer to the popular rough estimation method of the energy consumption of 5G networks as cited in the beginning of the paper as the baseline for comparison, which simply assumes that each 4G base station cell requires three 5G base station cells to achieve the same coverage based on their service coverage radiuses.

A summary of the results is shown in Table 2 (see Section B in Supplementary Information for more detailed description and illustrated maps), from which we can see that the spatial optimization does yield energy-saving effects. Particularly, 1) for the 4G scenarios, while the energy-saving effects when holding the number of base stations unchanged from the status-quo is mediocre (~2.5% - 5%), which is mostly gained from the reduced average distance between the base stations and traffic hotspots, the effects when both the number and locations of the base stations are optimized is considerablely large at ~ 20%-25%, a prominent margin over what the temporal shutdown approach would gain thanks to the reduction of necessary base station numbers [15–17]; 2) traffic hotspot areas-based optimization achieves more than an order of magnitude improvement in computational efficiency at the cost of energy efficiency improvement loss of about 20%. Besides, an examination of the results shows that load at any base station does not exceed 60%, implying that the increased loads in certain base stations do not affect the user experience due to network congestion, thus supporting the feasibility of the optimization; 3) the 5G scenario yields even more energy-efficiency improvement as compared to the 4G ones, with almost 30% of energy savings from the baseline.

**Table 2 Summary of the MCN energy efficiency optimization results**

| Scenario No. | Network type (service radius) | Solution space | Number of base stations as decision variable (alongside locations of base stations) | Baseline of energy savings effect | Baseline power (W) | Optimized power (W) | Energy savings effect | Calculation time cost (s) |
|---|---|---|---|---|---|---|---|---|
| I | 4G (r=300m) | Entire study area | N | Status-quo | 998,600 | 950,029 | 4.86% | 176.31 |
| II | | | Y | | | 751,675 | 24.73% | 32.01 |
| III | | Traffic hotspot areas | N | Status-quo | 998,600 | 973,966 | 2.46% | 12.63 |
| IV | | | Y | | | 802,907 | 19.59% | 3.31 |
| V | 5G (r=150m) | Traffic hotspot areas | Y | 4G Linear extrapolation | 2,356,875 | 1,657,353 | 29.68% | 2581.88 |

## Discussion

We proposed in this paper an MCN energy-saving strategy, which, given the existing practice of temporal energy optimization of the MCN and inspired by the principle of spatio-temporal symmetry in geography [33,34], pivots on the spatial dimension for additional space of MCN energy-saving. We showed deductively and empricially that the spatial distribution of MCN traffic satisfies the sknewness, clustering, and temporal stability conditions, such that a spatial optimization aiming to match the supply and demand of MCN resources would yield potentially considerable energy efficiency gain that are several times better than the currently widely used temporal shutdown-based measures [37]. In monetary terms, solely in China,our estimated 20%-30% energy savings would be equivilent to 20-30 billion RMB (about 3-4.5 billion US dollars) energy cost reduction by the official estimation of the expected market size of the 5G MCN [35], or 10-15 million tonnes reduction of $CO_2$ emission [36], a strong support for the value of our findings.

Moreover, the spatial optimization approach has significant advantages from the operational point of view. This is because, unlike the temporal shutdown measures, which relies on the prediction of real-time, wide-area traffic, the spatial optimization approach is based on the stable pattern of traffic distribution. Therefore, it is a once-and-for-all task over a reasonable time span. Besides, technically, changes in the spatial coverage of MCN resources in 5G networks can be realized using the beam-forming technology [4] without the need for large-scale physical rearrangement of existing base stations, which enhances the strategy's feasibility.

It should be noted that the energy saving potentials in this paper are estimated in simplified conditions, and real-world issues such as building shading in urban environments may hinder the spectrum efficiency such that the service coverage radius of base stations does not always reach the theoretical maximum. The resulting need for additional base stations would make the actual energy saving benefit lower compared to our simplified calculations. Nevertheless, our findings also suggest further energy saving potential through supply-demand coupling of MCN resources at finer spatial scales, which is technically enabled by the 5G MCN's beam-forming capaicities [4]. Indeed, limited by the spatial resolution of the raw data, we actually do not know the ground truth of spatial distribution of traffic at a smaller spatial resolution than the cell level. The data used in this paper, as the telecom operator preprocessed it, distributed the cell-level traffic equally among all analysis units (the 50 * 50m grids) within its coverage. However, both the maximum entropy derivation and the empirical calculations in Table 1 would imply that the actual traffic distribution is still likely to be highly skewed at a very small scale, in which case less base stations may be needed to accommodate the smaller traffic hotspots. That is to say, we may have underestimated the energy saving potential by using the relatively coarse-grained traffic data. Although, we do not know to what extent the two errors can cancel out each other, which constitutes a major limitation that awaits future empirical evidence.

The empirical results In this paper show a power-law spatial distribution of traffic at the citywide scale, along with power-law or exponential distributions at smaller scales (Fig. 3a). Interestingly, the spatial distribution of the population itself, as the user of the telecom service, is also found to have a similar scaling pattern [21,22]; and at large scales, the power parameters of both population and traffic distributions appear to be close to 2 [38]. As the spatial distribution of traffic can be viewed as that of the population after certain behavioral mapping, the coincidence seems to imply that the mapping may have a simple form, which is an interesting direction of future investigation.

Another interesting finding is the heterogeneity of the spatial distribution forms of traffic. Our findings show that at the same spatial scale, the sptial distribution of traffic in different areas of the city exhibit different forms (Fig. 3b, c). We speculate that the heterogeneity essentially derives from the same characteristics of the spatial distribution of mobile users, which in turn are related to the spatial heterogeneity of the city's functions and structures. Specifically, various parts of urban areas, or urban places at different scales, have different functional combinations and spatial forms. As different urban places may have different activating or inhibiting effects on people's traffic usage, the spatial distribution of traffic should reasonably have diversified forms. Nevertheless, if some small-scale urban places happen to have an mixture of urban functions and forms that are similar to those of the entire city, their traffic distribution may also exhibit similar characteristics to those of the city as a whole. This conjecture seems to explain the empirical findings in Table 1. While the revelation of the skewed spatial distribution pattern of traffic in this paper is sufficient to establish the basis for the strategy of achieving MCN energy savings through spatial optimization, it is still of great interest to reveal the root of this form of distribution. From the above discussion, it appears that the phenomenon is essentially a manifestation of the classical Modifiable Areal Unit Problem (MAUP) [25] in geography. Therefore, a deeper investigation of this problem may need to be supported by additional findings from urban science, as well as studies on the behavioral patterns of human mobility and traffic usage.

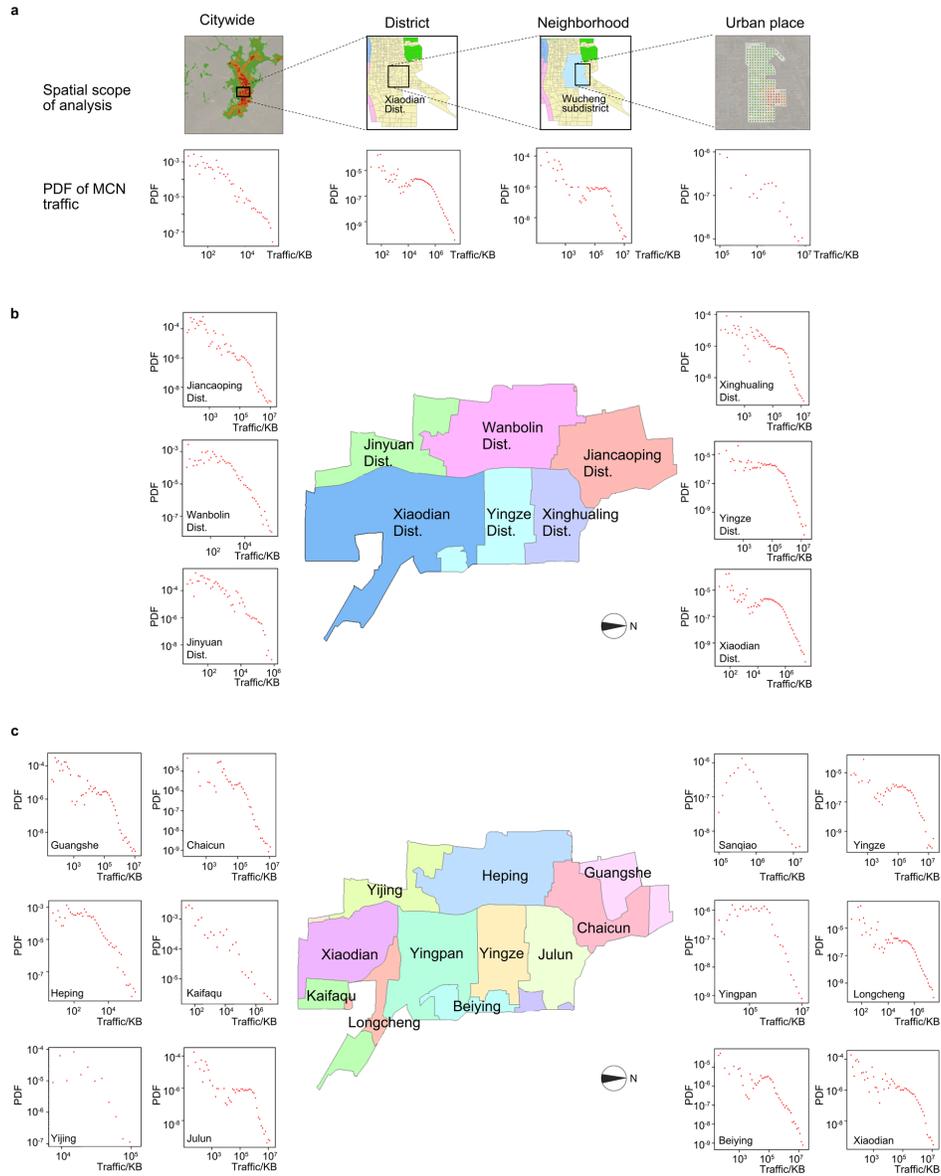

**Fig. 3 The scaling and spatial heterogeneous characteristics in the spatial distribution of traffic**

*a The spatial scaling patterns of the probability density distribution of traffic;*

*b The spatial heterogeneity patterns of the probability density distribution of traffic at the jurisdictional district level;*

*c The spatial heterogeneity patterns of the probability density distribution of traffic at the jurisdictional subdivision level.*

Finally, in an operational sense, the findings of this paper on the persistence of stable traffic hotspot areas hint at the potential advantages of using local rather than global solution in energy-efficient optimization of the MCN. Experiments show that the local approach with only the traffic hotspot areas as the solution space obtains a computational efficiency improvement higher than one order of magnitude with less than 20% loss of optimization benefit compared to the global solution approach, and thus has a significant advantage in terms of cost-efficiency. Considering the high requirement for agility in real-world energy-efficient optimization of the MCN [39], the result highlights the important operational significance of traffic hotspot area identification in practice.

# Methods

## Maximum entropy-based derivation of the spatial distribution form of traffic

By the maximum entropy assumption, the spatial distribution function $P(X)$ of the traffic should satisfy the following conditions:

$$\begin{aligned} min \quad & \sum_k P(X=k) \cdot \ln P(X=k) \\ s.t. \quad & \sum_k P(X=k) = 1 \\ & \sum_k P(X=k) \cdot c(k) = a \end{aligned} \quad (3)$$

Where $P(X=k)$ is the probability when the random variable $X$ for the value of the traffic (in bits) takes the value $k$. The first of the two constraints is a straightforward aggregate constraint. The second constraint introduces the traffic affine function $c(k)$, which gives the number of people (users) in the spatial location where $X=k$. Referring to previous studies on the relationship between traffic and population[20], we can reasonably assume that such mapping relationship exists. Therefore, the second constraint implies that the expectation of the population distribution in the city is convergent, which has been tested by previous studies [40].

Constructing the Lagrangian function for the solution, we obtain:

$$P(X=k) = e^{\lambda_0 - 1 + \lambda_1 \cdot c(k)} \quad (4)$$

Where the specific form of $c(k)$ is unknown. Nevertheless, based on findings in the urban science on the spatial distribution pattern of the urban population [40,41], and considering the heterogeneity of the urban space[42–44], we can conjecture several possible forms of $c(k)$. Depending on its skewness, we assume that $c(k)$ is logarithmically, linearly, power-law, Poisson, exponentially, and Gaussian distributed (see Section C in Supplementary Information for proof of convergence), and the respective solutions are:

$$P(X=k) = \begin{cases} e^{\lambda_0-1}k^{\lambda_1}; c(k) = \ln k \\ e^{\lambda_0-1}e^{\lambda_1 \cdot k}; c(k) = ak+b \\ e^{\lambda_0-1+\lambda_1 k^{-w}}; c(k) = k^{-w} \\ e^{\lambda_0-1+\lambda_1 e^{-t}\frac{t^k}{k!}}; c(k) = e^{-t}\frac{t^k}{k!} \\ e^{\lambda_0-1+\lambda_1 \cdot e^{-k}}; c(k) = e^{-k} \\ e^{\lambda_0-1+\lambda_1\frac{1}{\sqrt{2\pi}\sigma}e^{-\frac{k^2}{2\sigma^2}}}; c(k) = \frac{1}{\sqrt{2\pi}\sigma}e^{-\frac{(k-\mu)^2}{2\sigma^2}} \end{cases} \quad (5)$$

It can be seen that the form of $P(X = k)$ in the first case is a power-law distribution, and is exponential in the second; those in other cases are complex in form, but they can all be considered certain variations of generalized exponential distributions. Clearly, as the skewness of $c(k)$ increases, $P(X = k)$ changes from a power-law distribution to other forms of probability density distribution, and the skewness of all these forms are much higher than that of the power-law distribution. This means that among the possible forms of $c(k)$, the corresponding $P(X = k)$ distribution is at least power-law and may take on a more long-tailed distribution form (see Section A in Supplementary Information for an illustration of the skewness of the possible distributional forms). In terms of the energy saving potential incubated in such form of traffic distribution, a power-law form is "good" enough, while the others breeds much larger energy saving potential.

## MCN base station energy saving optimization problem statement

The base station in a 4G cellular network is an access link between the core network and the mobile devices (users). A base station consists of a set of equipments including power amplifiers, baseband units, RF units, power supplies, and air conditioning. The power of a base station in operation in a 4G cellular network $P_{op}$ is [32].

$$P_{op} = N_{TRX} \times [P_{PA}^{DC} + P_{RF}^{DC} + P_{BB}^{DC}]/(1-\sigma_{DC})(1-\sigma_{cool}) \quad (6)$$

Where $N_{TRX}$ denotes the number of transceivers in the base station; $P_{PA}^{DC}$ denotes the power of the power amplifier module; $P_{RF}^{DC}$ denotes the power of the RF unit; $P_{BB}^{DC}$ denotes the power of the baseband unit; and $\sigma_{DC}$ and $\sigma_{cool}$ denote the power loss of DC boost and air conditioning cooling, respectively, which take the values of 6% and 10%, respectively by general industry standards.

$P_{PA}^{DC}$ is a linear function of the base station transmit power $P_{tx}^{max}$. Denoting the amplifier efficiency with $\eta_{PA}$, we have:

$$P_{PA}^{DC} = P_{tx}^{max}/\eta_{PA} \qquad (7)$$

In general, both coverage and signal propagation attenuation are considered to be the main factors in determining the transmitting power of a base station. A simplified formula is expressed as follows:

$$P_{tx}^{max} = P_0(R/R_0)^\alpha \qquad (8)$$

Where $\alpha$ denotes the signal propagation path attenuation coefficient; $R_0$ denotes the standardized coverage radius of the base station with the standardized radius of 1 km; $P_0$ denotes the standardized power of the base station with the standardized power of 40 W.

Lastly, in addition to the operational power $P_{op}$, the base station includes a microwave antenna connecting the base station to the core network which has the power of $P_{mc}$, and a lighting module with the power of $P_{lm}$ [32]. Thus, the total power of an LTE-macro base station is:

$$P_{BS} = \frac{N_{TRX} \times \left(\frac{P_0(\frac{R}{R_0})^\alpha}{\eta_{PA}} + P_{RF}^{DC} + P_{BB}^{DC}\right)}{(1-\sigma_{DC})(1-\sigma_{cool})} + P_{mc} + P_{lm} \qquad (9)$$

Hence, holding operational time and the number of base stations constant, the optimization problem of base station distribution for minimizing energy consumption can be stated as follows:

$$\min: \sum_i P_i \tag{10}$$

$$s.t.\ SINR_k \geq SINR_{th}, \forall k = 1,2,\ldots,N_u; \tag{11}$$

$$P_i \leq P_{max} \tag{12}$$

Where:

$$\begin{aligned} SINR_k &= \frac{P_{re,k}}{I_{out,k} + \delta^2} \\ P_{re,k} &= \sum_{m=1}^{S} P_{b_{m,k}} g_{b_{m,k},k} \\ I_{out,k} &= \sum_{i=1, i \neq b_{m,k}}^{N_b} P_i g_{i,k} \\ g_{i,k} &= (D_{i,k})^{-a} L_0 \\ P_i &= \sum_k P_{i,k} \\ P_{i,k} &= P_{BBU} + P_{RRU_0} + \beta T_{i,k} \\ b_{m,k} &= \{i | D_{b_{m,k}} = \min\{D_{1,k}, D_{2,k}, \ldots, D_{N_b,k}\}\} \end{aligned} \tag{13}$$

Where $P_i$ denotes the power of base station $i$. The two constraint terms include the signal strength constraint and the power rating constraint. Specifically, for the first constraint, $SINR$ denotes the signal-to-noise ratio at user $k$, which is equal to the ratio of the received base station transmit power at user $k$ ($P_{re,k}$) to the sum of the received interference power at user $k$ ($I_{out,k}$) and the ambient noise $\delta^2$, where $P_{re,k}$ is equal to the product of the transmit power of the closest base station at user $k$ ($P_{b_{m,k}}$) and the transmit loss ($g_{b_{m,k},k}$), the latter of which is related to the transmit distance $D_{i,k}$ and the path loss factor $L_0$; and $I_{out,k}$ denotes weak signal interference base station power received at user $k$. For the second constraint, $P_i$ is the sum of $P_{i,k}$ at all locations $k$, which is the power consumed by base station $i$ for the traffic provided to location $k$, and can be computed as the sum of the baseband unit power ($P_{BBU}$), the RF unit base power ($P_{RRU_0}$), and the power for transmitting data packets $\beta T_{i,k}$. The coefficient $\beta$ describes the relationship between traffic volume and the respective power [12,13]. And $P_{max}$ is the power rating of the base station. Note that we did not explicitly include the load of base stations in the optimization problem, as in real-world MCN operations the factor rarely exceeds 30%, and thus does not affect the user experience of MCN services. However, we do validate that the factor does not exceeds a reasonable level in our results.

When discretizing the solution space with a 50 * 50m grid, and holding the number of base stations as is in the status-quo, the above optimization problem can be transformed into an integer programming problem:

$$\max \sum_{i \in I} y_i \quad (14)$$

$$s.t. y_i \in \{0, 1\}, i \in I \quad (15)$$

$$x_j \in \{0, 1\}, j \in J \quad (16)$$

$$\sum_{j \in J} x_j = N_b \quad (17)$$

$$\sum_{j \in N_i} x_j \geq y_i, N_i = j \in J : d_{i,j} \leq R \quad (18)$$

Where $y_i$ denotes the required coverage locations in the location space $I$, and $\{0, 1\}$ represents whether a location is covered; $x_j$ denotes possbile base station locations in the location space $J$, and $\{0, 1\}$ represents whether a location is chosen. $N_b$ denotes the number of base stations in the result. $d_{i,j}$ denotes the distance between $i$ and $j$. And $R$ represents the base station coverage area.

Further, when the number of base stations ($N_b$) is also optimized, the original optimization objective is stated as (with the constraints unchanged):

$$min: \begin{cases} \sum_i P_i \\ N_b \end{cases} \quad (19)$$

We now have to filter out the unnecessary base stations. And the integer programming problem now can be stated as:

$$min \sum_{i \in I} x_i \quad (20)$$

$$s.t. x_i \in \{0, 1\}, i \in I \quad (21)$$

$$\sum_{j \in N_{i,j}} x_i \cdot y_j \leq C, N_{i,j} = \{j | d_{i,j} \leq d_{p,j}, for\ p\ in\ I\} \quad (22)$$

$$\sum_{i \in I} x_i \cdot D_{i,j} \geq 1, for\ j \in J \quad (23)$$

$$D_{i,j} \begin{cases} 0, d_{i,j} > R \\ 1, d_{i,j} \leq R \end{cases} \quad (24)$$

The additional constraints are necessary because we have to keep the traffic load of a base station under its capacity, which is directly related to the service coverage radius $R$ of the base station. The specific value of $R$ is related to the terrain status of the service area, and in this paper we take a 300m value for the 4G network and 150m for the 5G network, which are drawn from experience [45].

## Data

The MCN traffic data used in this paper are for the urban area of Taiyuan, the capital of Shanxi Province, China (a map of the study area is available in Section D in Supplementary Information), from January 7, 2020 to January 17, 2020, with an hourly temporal resolution and cell-level spatial resolution. The original data includes information on upstream and downstream traffic values, number of users, and MCN signal strength in each cell. The data has been resampled to have a 50 * 50m spatial resolution by the data provider (the telecom operator) through evenly distributing the traffic of a cell among its coverage area.

## Data Availability

The original data of MCN traffic used in this paper is temporarily not publicly available by data provision terms from the data provider.

## Code Availability

The model and code used in this study are available upon request.

## Ethics declarations

Competing interests

The authors declare no competing interest.